\documentclass[12pt]{article}
\usepackage{amsmath}
\usepackage{amsthm}
\usepackage{amsfonts}
\newcommand{\bZ}{{\mathbb Z}}
\newcommand{\dS}{{\mathbb S^2}}
\newcommand{\tT}{{\mathbb T^3}}
\newtheorem{theorem}{Theorem}
\newtheorem{lemma}{Lemma}
\newtheorem{corollary}{Corollary}
\begin{document}

\begin{center}
{\Large \bf
Existence and measure of ergodic leaves in Novikov's problem
on the semiclassical motion of an electron
}
\\[20pt]
Roberto De Leo\\
rdl@math.umd.edu\\
Department of Mathematics\\
University of Maryland\\
College Park, MD 20742, USA\\[15pt]
\end{center}
\begin{abstract}
\sl
We show that ``ergodic r\'egime'' appears for generic dispersion relations in the
semiclassical motion of electrons in a metal and we prove that, in the fixed energy
picture, the measure of the set of such directions is zero.
\end{abstract}

The problem of semiclassical motion of an electron in a lattice under a strong magnetic
field leads to a very interesting problem of low dimensional ``periodic'' topology \cite{Nov82},
namely the structure of plane sections of 2-dimensional submanifolds of $\tT$.
A.V. Zorich \cite{Zor84} and I.A. Dynnikov \cite{Dyn93a} proved Novikov's conjecture, i.e. 
that open orbits are strongly asymptotics to some straight line, respectively in the almost 
rational and generic case; from their analysis Novikov \cite{Nov95} extracted the following 
beautiful picture: once a dispersion relation is given, to every magnetic field direction 
giving rise to the generic situation (in which open leaves fill a genus-1 component of the 
Fermi surface) it corresponds one and only one indivisible integer 2-cycle in $\tT$ 
(Miller index). Such triple of integers are locally constant with respect to the direction of
magnetic field.\par
From this construction we get two kinds of pictures on the 2-sphere. One is the so-called
{\sl global} picture, in which we label a point with the relative integer 2-cycle if it 
gives rise to the generic situation for some value of energy. In this case on the 2-sphere
it is defined a countable set of disjoint open sets whose union is dense. It is still unclear 
whether this set has full measure or not.\par
The second kind of pictures is the {\sl fixed energy} one, in which fix a value of the energy 
and we label a point with the corresponding integer 2-cycle only if it gives rise to generic
open orbits at that energy, and we label a point with the null 2-cycle if all orbits are
closed at the same energy.\par
Novikov and Maltsev \cite{NM98} applied these results to the conductivity theory of normal
metals describing new universal observable topological phenomena.
It is hence very important, also from the physical point of view, to understand more about the
appearance of these special directions of the magnetic field.\par
Below we present a series of
statements that represent the stronger properties known at this time about their measure.
These theorems are needed for applications in physics.
They are based on the techniques developed by Dynnikov and were missing in the previous 
publications. As I was informed in process of publication of this article, Dynnikov also 
published some of these results in his new article \cite{Dyn99}.

\begin{lemma}
For a generic dispersion relation $f\in C^\infty(\tT)$, boundaries of stabilities zones 
are piecewise smooth and stability zones meet in a countable set of points.
\end{lemma}

\begin{corollary}
For every generic dispersion relation $f\in C^\infty(\tT)$ there are uncountably
many ergodic directions.
\end{corollary}

\begin{theorem}
For any $f\in C^\infty(\tT)$, the set of directions generating ``ergodic r\'egime'' 
in the fixed energy picture has zero measure for almost all values of $f$.
\end{theorem}

\begin{theorem}
For a generic $f\in C^\infty(\tT)$, the set of directions generating ``ergodic r\'egime'' 
in the fixed energy picture has zero measure for all values of $f$.
\end{theorem}

The idea that leads to the corollary is that on every generic loop in $\dS$ the set
of non generic directions forms a Cantor set, so that the set of all non generic 
directions forms some kind of fractal on the sphere. \par
An analytical study of the properties of this fractal appears to be very difficult
to perform. It is know from theorems 1 and 2 that for a generic oriented surface
the set of non generic direction has measure 0, but it is still an open question even
whether this measure remains zero when we consider the set of ergodic direction relative 
to an interval of energies, that is relevant for \cite{NM98}.\par
It has been conjectured by S.P. Novikov that generically the set of ``ergodic'' directions
at fixed energy has fractal dimension $\alpha\in(0,1)$ on the sphere, and similarly that the
set of all of them should have fractal dimension $\beta\in(1,2)$.\par
Numerical computations have been performed by the author \cite{DL99} for the function
$$f(x,y,z) = \cos(x) +  \cos(y) + \cos(z)\;\;.$$
Because of the symmetry $f(x^\alpha+\pi)=-f(x^\alpha)$, this function has the property to
give rise to all stability zones at their biggest size at the same energy, namely at $E=0$,
so that the global picture can be obtained just studying a fixed energy picture.\par
The numerical simulation works in the following way:
the surface $M^2 = f^{-1}(0)$ has genus 3
and the embedding $i:M^2\to\tT$ has rank 3, so that we can find a canonical base of cycles
$\{e_j,f_j\}$ on $M^2$ s.t. $i_*(e_j)=0$ and $i_*(f_j)$ are a base in $H_1(\tT,\bZ)$.
The program scans then all rational values of the magnetic field $H=(m/N,n/N,1)$,
$N\geq n\geq m$,
for some fixed $N$, and for any of them finds the critical points and evaluates the homology
class of the two critical loops in $\tT$ and its intersection numbers with the cycles
$i_*(f_j)$ in $M^2$.
If just one of the critical loops is homologous to zero in $\tT$, then the intersection
number of the other gives automatically the homology class of the indivisible 2-cycle
$l\in H_2(\tT,\bZ)$ that corresponds to that magnetic field.\par
The following table shows the results obtained for $N=400$:

$$
\vbox{\tabskip=0pt \offinterlineskip
\def\tablerule{\noalign{\hrule}}
\halign{
  \strut#&\vrule#\tabskip=1em plus2em&
  \hfil$#$\hfil&\vrule#&\hfil$#$\hfil&\vrule#&
  \hfil$#$\hfil&\vrule#&\hfil$#$&\vrule#\tabskip=0pt\cr\tablerule
&&\omit\hidewidth Hom Class\hidewidth&&
 \omit\hidewidth Area\hidewidth&&\omit\hidewidth Hom Class\hidewidth&&
 \omit\hidewidth Area\hidewidth&\cr\tablerule
&&(0,0,1)&&(2.83\pm.02)10^{-1}&&(1,5,5)&&(4.1\pm.2)10^{-3}&\cr\tablerule
&&(1,1,1)&&(2.03\pm.01)10^{-1}&&(2,5,8)&&(4.1\pm.4)10^{-3}&\cr\tablerule
&&(1,2,2)&&(8.2\pm.2)10^{-2}&&(2,6,7)&&(3.4\pm.4)10^{-3}&\cr\tablerule
&&(0,1,2)&&(5.1\pm.1)10^{-2}&&(4,7,8)&&(3.0\pm.3)10^{-3}&\cr\tablerule
&&(1,3,3)&&(2.1\pm.1)10^{-2}&&(0,3,4)&&(2.9\pm.4)10^{-3}&\cr\tablerule
&&(2,3,4)&&(1.7\pm.1)10^{-2}&&(3,5,7)&&(2.7\pm.3)10^{-3}&\cr\tablerule
&&(1,3,5)&&(9.6\pm.5)10^{-3}&&(1,6,6)&&(2.0\pm.1)10^{-3}&\cr\tablerule
&&(1,4,6)&&(9.6\pm.5)10^{-3}&&(4,5,8)&&(2.0\pm.4)10^{-3}&\cr\tablerule
&&(0,2,3)&&(9.0\pm.6)10^{-3}&&(5,8,10)&&(1.9\pm.4)10^{-3}&\cr\tablerule
&&(2,4,5)&&(8.6\pm.6)10^{-3}&&(4,6,9)&&(1.8\pm.3)10^{-3}&\cr\tablerule
&&(1,4,4)&&(8.3\pm.3)10^{-3}&&(1,6,10)&&(1.7\pm.1)10^{-3}&\cr\tablerule
&&(1,2,4)&&(6.2\pm.5)10^{-3}&&(5,9,11)&&(1.6\pm.2)10^{-3}&\cr\tablerule
&&(3,4,6)&&(4.7\pm.5)10^{-3}&&(4,6,7)&&(1.5\pm.2)10^{-3}&\cr\tablerule
}}
$$

We used two different standard techniques to evaluate the Minkowski fractal 
dimension of the set of ``ergodic'' directions from the data we obtained 
puttin $N=400$. The values obtained for the fractal dimension agree very
well with each other and give an estimate of $d\simeq1.8$, consistently with 
Novikov's conjecture.
New calculations for $N=1000$ are on the way so that it will be possible to 
have a better estimate in a short time.
\vfill\eject
\vskip 1.cm \noindent{\bf Acknowledgments}\par The author gratefully
thanks his advisor S.P. Novikov for introducing the subject and for
many useful advices and suggestions, and also thanks I.A. Dynnikov and
A. Giacobbe for many helpful discussions. The author
also acknowledge financial support from Indam for its second and third
year of PhD at UMD.

\end{document}